\documentclass[a4paper,11pt]{article}
\pdfoutput=1
\usepackage{jheppub}
\usepackage[T1]{fontenc}
\usepackage{indentfirst}
\usepackage{setspace}
\usepackage{color}
\usepackage{graphicx}
\usepackage{url}
\usepackage{dsfont}
\usepackage{bookmark}
\usepackage{soul}
\usepackage{pdfpages}

\usepackage{amsmath, amsfonts, amssymb}
\usepackage{mathtools}
\usepackage{physics}
\usepackage{mathrsfs}
\usepackage{tensor}

\usepackage{hyperref}
\usepackage{natbib}
\usepackage{footnote}
\usepackage{tablefootnote}

\title{Dark Energy as a Post-Inflation Effect in Quadratic Gravity}
\author[a]{Heng-Wei Chang}
\affiliation[a]{Department of Physics, National Tsing-Hua University,
  Hsinchu 30013, Taiwan}
\date{\today}
\emailAdd{wayne86122488@gapp.nthu.edu.tw}

\abstract{
	We analytically and numerically show that the acceleration of the cosmic expansion could be explained 
	by a Quadratic Gravity model which is known to be able to trigger sufficient inflation, 
	with neither negative pressure matter nor cosmological constant.
	Furthermore, we exactly show that this model differs greatly from Einstein's gravity in radiation and matter dominant era,
	giving a non-shrinking Hubble horizon and a smooth expansion history of the universe.
	Accordingly, it suggests that the Dark Energy could possibly be a post-inflation effect in Quadratic Gravity. 
	We also show that this model admits all Einstein metrics as its solutions. 
	Consequently, classic tests of Einstein’s Gravity cannot falsify this model.
}

\keywords{Inflation, Dark Energy, Modified Gravity}

\begin{document}

\setlength{\parskip}{0.6em}
\setstretch{1.6} 
\maketitle
\setcounter{page}{1} 
\setstretch{1}

\section{Introduction}

Quadratic Gravity theories have many inspiring properties. 
Firstly, they’re commonly renormalizable for their modified graviton propagators \citep{PhysRevD.16.953, Salvio_2018}. 
Secondly, Conformal Gravity, one special case of it, 
contains spherically symmetric solutions which fit the galaxy rotation curve well \citep{Mannheim:1988dj, Mannheim:2011ds}. 
Another example is the Starobinsky model, 
which can provide slow-roll and 60 e-folded expansion, causing Inflation to resolve the fine-tuned problems
\citep{Linde:1983gd, Liddle:1999mq, liddle:2000cg, Guth2007, Linde_inflation, 1980PhLB...91...99S}.

Hence, it's then natural to ask whether quadratic gravity theories could explain the accelerating cosmic expansion
\citep{Li:2012dt, COPELAND_2006, PhysRevD.74.086009} without unphysical matter with negative pressure
in contrast to the quintessence model \citep{PhysRevD.37.3406, PhysRevD.59.123504},
which is the main goal of this paper.

However, Higher Derivative Gravity Theories (HDGT) do have their drawback.
For example, some of them, like Conformal Gravity, are not renormalizable \citep{Salvio_2018_2}.
Another common issue is the existence of ghost modes \citep{Niedermaier_2007,Mannheim21,N_ez_2005}, 
which drives us to regard HDGT only as a low energy effective field theory of 
a ghost-free and renormalizable (or even immune from renormalization) quantum gravity 
which works effecitvely below some energy scale merely at the classical level.
However, recent studies have proposed that we could possibly cure the ghost problem \citep{Anselmi_2017, Anselmi_2017_2, Strumia_2018, Salvio_2019, Salvio_2021}.

Also, this is not the first time there's the idea that Inflation and Dark Energy could be unified in a single model \citep{Nojiri_2003}.
But with the belief that the universe might once be at some conformal UV fixed point of a "grand unifying" theory right at the Big Bang,
flowing to a conformal symmetry broken phase and giving some length scales such as the size of the Hubble horizon,
we are interested whether some HDGT could effectively describe this at the classical level and resolve the unsolved problems in Einstein's gravity at the same time.

The simplest action satisfying these considerations is
\begin{equation}\label{eq:lagrangian}
	S = \int\sqrt{|g|} d^4x \Big(-\frac{\alpha}{2} R^2-\frac{\beta}{2} R_{\mu\nu}R^{\mu \nu}
	-\frac{1}{2}(\partial \varphi)^2 +\frac{\xi}{2}\varphi^2R+\frac{\nu}{4}\varphi^4 + \mathcal{L}_m\Big),
\end{equation}
where $\varphi$ is a real scalar field (possibly dilaton), 
$\alpha, \beta, \xi,$ and $\nu$ are real constants, 
and $\mathcal{L}_m$ is the Lagrangian of the other matter fields. 
For instance, different species of perfect fluids which appears later in our discussion.
A term such as the contraction of the Riemann curvature with itself could be added, but it can be canceled by the Gauss-Bonnet term. Another possible term is the Laplacian of the Ricci scalar, but it’s also a total derivative.

The action \eqref{eq:lagrangian} can describe a conformally broken phase
as it's known to be conformally invariant at the classical level only if $3\alpha+\beta = 0$ and $1+6\xi = 0$.
For convenience, we call this action Quadratic Gravity theory with broken conformal symmetry if the both of the equalities do not hold.

However, any pratical gravity model shall still pass the classic tests of the Einstein's gravity,
including the precession of the Mercury and the deflection of light by massive compact stars, 
both of which are based on the Schwarzschild solution.
This holds in this model. In fact, in the dimension of four, 
vacuum solutions of the Einstein equations solve the equations of motion of the quadratic gravity part of this model,
\begin{equation}
W_{\mu \nu} =0.
\end{equation}

Here  $W_{\mu\nu} = W^1_{\mu\nu} - W^2_{\mu\nu}$ is the generalized Einstein tensor, with
\begin{equation} \label{eq:W1}
	W^1_{\mu\nu} 
	= \alpha \nabla_\mu \nabla_\nu R 
	- (\alpha +\frac{\beta}{4}) g_{\mu\nu} \Box R
	+ \beta  (\nabla_\nu \nabla_k \tensor{R}{^k_\mu}
	- \frac{1}{2}\Box R_{\mu\nu})
\end{equation}
and
\begin{equation} \label{eq:W2}
	W^2_{\mu\nu} = \alpha R(R_{\mu\nu}-\frac{1}{4} g_{\mu \nu} R)
	+ \beta  (\tensor{R}{_\mu^k} R_{k\nu} - \frac{1}{4} g_{\mu\nu}R_{ab}R^{ab}).
\end{equation}

For an Einstein manifold
$R_{\mu\nu} = \Lambda g_{\mu\nu}$, $W^1 = 0$
due to the metric compatibility of the Levi-Civita connection.
One also can see that $W^2$ vanishes since  
$W^2_{\mu\nu} = \Lambda^2 (\tensor{\delta}{^\kappa_\kappa} \alpha +  \beta) (1-\frac{1}{4} \tensor{\delta}{^\sigma_\sigma}) g_{\mu\nu} = 0$.
On the other hand, the scalar field equation of motion is given by
\begin{equation}
	\Box \varphi + \xi R\varphi + \nu \varphi^3 = 0,
\end{equation}
and its energy momentum tensor is
\begin{equation}
	S_{\mu\nu} = \frac{1}{2} \partial_\mu \varphi \partial_\nu \varphi 
	- \frac{1}{2} g_{\mu\nu} \Big( \frac{1}{2}(\partial \varphi)^2 -\frac{\nu}{4} \varphi^4 \Big)
	- \frac{\xi}{2}\Big(g_{\mu\nu} \Box( \varphi^2) - \nabla_\mu \nabla_\nu \varphi^2 + G_{\mu \nu} \varphi^2\Big),
\end{equation}
where $G_{\mu\nu}$ is the Einstein tensor and $\Box = \nabla^\mu \nabla_\mu$ is the d'Alembert operator.
If the spacetime is Einstein and we take $\varphi^2 = -4\xi\Lambda/\nu$, 
then the generalized Einstein equations are satisfied
since $W_{\mu\nu} = 0 = S_{\mu\nu}$. 
The scalar field equation of motion is also clearly satisfied since it's at the minimum of the potential.  
Hence, all Einstein manifolds are allowed.

In fact, this is not the first time this action is considered.  
In \citep{Rinaldi_2016, Vicentini_2019}, 
it's shown that this model can satisfy the inflation requirements for its Starobinsky like nature.  
Consequently, what left to study is whether this model can explain the galaxy rotational curves,
which is difficult due to the complexity of the equations of motion on general spherically symmetric metrics,
and the acceleration of the cosmic expansion, which is our main goal.

To achieve our ambition, 
we'll simplify the cosmological equations of motion and discuss the integrability of this model in the first section.

In the second part,
presence of single species of perfect fluid is firstly considered, 
and we construct an exact solution to show that this model allows an smooth transition
from power law expansions to an exponential one in the sparse density limit, i.e. in the vacuum.
Since the equations of motion are unlikely solvable with multiple species of perfect fluids,
numerical analysis is then done to fit the model with the type Ia supernova data. 

In this paper, we implement the convention $R_{ijmn} = \Gamma^p_{in} \Gamma_{jmp} - \Gamma^p_{im}\Gamma_{pjn} + \partial_n\Gamma_{jim} - \partial_m\Gamma_{jin}$
for the Riemann curvature tensor. Also, $\log$ always refers to the natural logarithm.

\section{The Cosmological Equations of Motion and the Analytical Properties}

Let us consider the Friedmann–Lemaître–Robertson–Walker metric
\begin{equation} \label{eq:flrw_metric}
	ds^2 = - dt^2 + a(t)^2\Big( \frac{dr^2}{1-kr^2} + r^2 d\Omega_{S^2} \Big),
\end{equation}
where $d\Omega_{S^2}$ is the metric of a unit 2-sphere,
$a(t)$ is the cosmological evolution scale factor, 
and $k$ is the spatial curvature, which determines the ``shape'' of the universe. 
For brevity, we'll write $a(t)=e^{b(t)}$.
Also, in this paper we will only consider $\varphi = \varphi(t)$ for classical cosmology.
Because of the homogeneity and the isotropicity of the metric,
the tensors $W^\mu_\nu$ and $S^\mu_\nu$ are diagonal, 
and their three spatial components are identical respectively. 
Hence, if we don't consider the presence of other matter, 
the equations which we need solve are the scalar field equation of motion, 
$W^t_t = S^t_t$, and the trace equality $\tensor{W}{^\mu_\mu}=\tensor{S}{^\mu_\mu}$.
However, vanishing divergence of these tensor can both give us 
$ \frac{d}{dt}(a^4\tensor{W}{^t_t}) = \dot{a} a^3\tensor{W}{^\mu_\mu}$ 
and $ \frac{d}{dt}(a^{4}\tensor{S}{^t_t}) = \dot{a} a^{3}\tensor{S}{^\mu_\mu}$.
As a result, we only need to solve the scalar field equation of motion
\begin{equation}
	\ddot{\varphi} + 3\dot{b} \dot{\varphi} - \nu \varphi^3 - 6\xi\varphi(\ddot{b}+2\dot{b}^2+k e^{-2b}) = 0
\end{equation}
and $\tensor{W}{^t_t} = \tensor{S}{^t_t}$, where
\begin{equation}
	\tensor{W}{^t_t} = 3(3\alpha+\beta)(2\dot{b}\dddot{b} + 6 \dot{b}^2 \ddot{b} - \ddot{b}^2 -2 k e^{-2b} \dot{b}^2 + k^2 e^{-4b})
\end{equation}
and
\begin{equation}
	\tensor{S}{^t_t} = \frac{1}{8} \Big( \nu \varphi^4 + 12 \xi \varphi^2 (\dot{b}^2 + k e^{-2b}) +24 \xi\dot{b} \varphi \dot{\varphi} -2 \dot{\varphi}^2 \Big).
\end{equation}

To simplify these equations, consider the conformal time scale $\tau := \tau_0+\int^t_{t_0} a(t')^{-1}dt'$
and the field redefinition $\psi(\tau) := a(\tau)\varphi(\tau).$
In this way, the equations are equivalent to 
\begin{equation} \label{eq:einstein_eq}
	\begin{split}
	24(3\alpha+\beta)&(2\eta\eta''-\eta'^2-3\eta^4-2k\eta^2+k^2) \\
	&= \nu \psi^4 +12k\xi\psi^2+ 2(1+6\xi)(2\psi'-\eta\psi)\eta\psi -2\psi'^2 
\end{split}
\end{equation}
and
\begin{equation} \label{eq:scalar_eom}
	\psi'' -\Big( 6k\xi + (1+6\xi)(\eta^2+\eta')\Big)\psi-\nu \psi^3 =0.
\end{equation}
Here, we use $'$ to represent the differentiation with respect to $\tau$ and denote $\eta = b'$ for simplicity.
Hence, we find a second order expression for the equations of motion, which is more convenient for both analytic and numerical studies.

\subsection{The Partial Integrability and the Accidental Integrability}

We accidentally found that the equation \eqref{eq:einstein_eq} can be ``integrated'' to
\begin{equation} \label{eq:ansatz}
	6(3\alpha+\beta)\Big( \eta'^2 - (\eta^2+k)^2 -\gamma \eta \Big) =
	\frac{1}{2}\psi'^2-3k\xi \psi^2 -\frac{\nu}{4} \psi^4 - \frac{1+6\xi}{2}\eta^2\psi^2,
\end{equation}
where $\gamma$ is some constant.
Furthermore, if we denote the right hand side of this equation by $\sigma$, then by \eqref{eq:scalar_eom} we have
\begin{equation} \label{eq:sigma_evo}
	\sigma' = (1+6\xi)(\psi'-\eta \psi)\psi \eta'.
\end{equation}
By definition, that $\psi' = \eta \psi$ is equivalent to that $\varphi$ is a constant.
As a result, this system is completely integrable if $1+6\xi=0$, $\varphi$ is a constant, or $\eta$ is a constant.

What is the origin of this integrability?
Are they Noether charges of some continuous symmetries?
Since this is an autonomous system of differential equations, one then may speculate that the first integrability would be the Hamiltonian. 
To verify this, we firstly calculate the action with our change of variables, and it's
\begin{equation} \label{eq:effective_lagrangian}
\begin{split}
	S &= \int \frac{4\pi r^2}{1-kr^2}dr \int d\tau \Big( -6(3\alpha+\beta)(k^2+\eta^4+\eta'^2 + 2k \eta^2 )  \\ 
	  & \hspace{4em}-6 (6\alpha+\beta)(k\eta'+ \eta^2\eta') + \frac{1}{2}(\psi'-\eta\psi)^2 + 3\xi (k + \eta^2 + \eta') \psi^2 + \frac{\nu}{4} \psi^4 \Big) \\
	  &:= \int \frac{4\pi r^2}{1-kr^2} dr \int d\tau \mathcal{L}_\tau.
\end{split}
\end{equation}
Also, Legendre transformation then gives the Hamiltonian density 
\begin{equation}
	\begin{split}
	\mathcal{H}_\tau &:= \frac{\partial \mathcal{L}_\tau}{\partial \eta'} \eta' + \frac{\partial \mathcal{L}_\tau}{\partial \psi'} \psi'  - \mathcal{L}_\tau \\ 
	&= 6(3\alpha+\beta)\Big((\eta^2+k)^2 - \eta'^2\Big) + \frac{1}{2}\psi'^2 - 3k\xi\psi^2 - \frac{\nu}{4}\psi^4 - \frac{1+6\xi}{2} \eta^2 \psi^2.
\end{split}
\end{equation}

This is clearly different from \eqref{eq:ansatz}. Furthermore, one can easily verify that this Hamiltonian is not even conserved by the equations of motion. 
We explain below why this ``Hamiltonian'' is not conserved.

As we conduct variation of the action with respect to the metric, what is varied is $a(\tau)$ instead of $\eta$,
although it does not appear explicitly in the action \eqref{eq:effective_lagrangian}.
In general, consider a theory does not depend on its dynamical variable $x$ but its derivatives $\dot{x}$ and $\ddot{x}$, i.e. Lagrangian $L=L(\dot{x},\ddot{x})$.
The Euler-Lagrange equations of this Lagrangian shall be 
\begin{equation}
	\frac{d^2}{dt^2}\Big( \frac{\partial L}{\partial \ddot{x}} \Big) - \frac{d}{dt}\Big( \frac{\partial L}{\partial \dot{x}} \Big) = 0,
\end{equation}
which can be integrated to 
\begin{equation}
	\frac{d}{dt}\Big( \frac{\partial L}{\partial \ddot{x}} \Big) -  \frac{\partial L}{\partial \dot{x}} = K.
\end{equation}
However, if we inappropriately treat $y = \dot{x}$ as the general coordinate and vary the action with respect to it, 
then the equation of motion will unreasonably require $Q =0$. Furthermore, the Hamiltonian calculated in this way is also not conserved since 
\begin{equation}
	\frac{d H_y}{d t} = \frac{d}{dt} \Big( \frac{\partial L}{\partial \dot{y}}\dot{y} - L\Big) 
	= \Big( \frac{d}{dt}\Big( \frac{\partial L}{\partial \ddot{x}} \Big) - \frac{\partial L}{\partial \dot{x}}  \Big) \dot{y} = K \dot{y}. 
\end{equation}
As a result, $H_y - K \dot{x}$ is the authentically conserved one.

On the other hand, $\sigma$ defined in \eqref{eq:ansatz} is a constant if $1+6\xi=0$, 
consequently, one may guess that it could be the conformal symmetry Noether charge.
Under conformal transformation $g \rightarrow \Omega^2g$, 
we have $a \rightarrow \Omega a$ and $\varphi \rightarrow \Omega^{-1}\varphi$, thus, $\psi \rightarrow \psi$. 
By the Noether theorem, we then have the conformal symmetry conserved charge
\begin{equation}
	Q = \frac{\partial L}{\partial \eta} - \frac{d}{d\tau} \Big( \frac{\partial L}{\partial \eta'}\Big), 
\end{equation}
but this is just the previous $K$.
Since we don't know what kind of symmetry could bring us this conserved charge,
we then call the conservation of $\sigma$ an ``accidental integrability'' in exactly the same sense of the Laplace-Runge-Lenz vector in the Newton gravity.

\subsection{The Exponentially Expanding Dark Energy Solutions}

The accidental integrability \eqref{eq:sigma_evo} could provide us with some trivial but interesting solutions.
For example, if the scalar field  $\varphi$ is a constant, or $\psi' = \eta\psi$, 
the equations of motion then imply $\nu\psi^2+12\xi(k+\eta^2)=0$ and $\eta' = k +\eta^2$,
which can be easily studied.
In fact, this exactly reduces to the Einstein gravity.
Hence, if we want to search for non-Einstein cosmological solutions, then we'd expect that $\varphi$ would not be merely a constant.

However, since quadratic gravity with broken conformal symmetry allows all Einstein solutions,
one would like to know what the dark energy solution in the $\Lambda$-CDM model would be in our notation.
If we consider $a(t) = e^{\eta_0(t-t_0)}$ and $\varphi = \psi_0$, then it turns out that this exponentially expanding solution is equivalent to
\begin{equation}
	\eta(\tau) = \frac{\eta_0}{1-\eta_0 \tau} 
	\text{ \hspace{1em} and \hspace{1em}  }
	\psi(\tau) = \frac{\psi_0}{1-\eta_0 \tau},
\end{equation}
which obeys $\eta' = \eta^2$, requiring $k=0$.
This expression of the dark energy solution will be important in later discussions, 
especially in the discussion of power transition solutions.

\section{Dark Energy as a Post-Inflation Effect of Inflation}

After discussing the analytic properties of the quadratic gravity theory with broken conformal symmetry,
we now turn to the empirical aspects of it.
Firstly, we will consider the presence of a single perfect fluid and
work out the cosmological history given by this model.
After this, we'll consider the presence of both cold dust and radiation, as the standard cosmology does,
and we then fit it with the Supernova data in \citep{2018_sharov}.

\subsection{Analytic Results with a Single Species of Perfect Fluid}

Inspired by the single perfect fluid exact solutions in Einstein's gravity \citep{liddle:2000cg}, we consider the ansatz
\begin{equation}
	a(\tau) = \Big(\frac{1}{1 - \eta_0 \tau}\Big)^p
\end{equation}
and
\begin{equation}
	\psi(\tau) = \psi_0 \Big(\frac{1}{1 - \eta_0\tau} \Big)^q,
\end{equation}
for some constant $\eta_0, \psi_0, p,$ and $q$.
In this case, $\eta$ is given by 
\begin{equation}
	\eta(\tau) = \frac{p\eta_0}{1-\eta_0 \tau}.
\end{equation}
If $p=1$, then it becomes the dark energy solution we mentioned before.
For a perfect fluid with density $\rho(\tau)$ and pressure $P(\tau)$, which obeys the equation of state $P = \omega \rho$,
the continuity equation implies that 
\begin{equation}
	\rho' = -3(1+\omega)\eta\rho.
\end{equation}
For convenience, we define $\tilde{\rho} = e^{4a}\rho$, then we have 
\begin{equation}
	\tilde{\rho}' = (1-3\omega)\eta\tilde{\rho},
\end{equation}
which implies
\begin{equation}
	\tilde{\rho}(\tau) = \tilde{\rho}(0)\Big(\frac{1}{1-\eta_0 \tau}\Big)^{p(1-3\omega)}.
\end{equation}
Substitute these into the equations of motion with $k = 0$ and implement the power matching, 
then one will get $q=1$ and $p(1-3\omega) = 4$, provided that $\omega\neq1/3$.
Hence, in this case we have 
\begin{equation}
	a(\tau) = \Big(\frac{1}{1-\eta_0 \tau}\Big)^{\frac{4}{1-3\omega}},
\end{equation}
with a relation
\begin{equation}
	\nu\psi^2_0 = \Big(2-(1+6\xi)p(p+1) \Big)\eta_0^2
\end{equation}
between these constants.

In the relativistic fluid case $\omega = 1/3$, 
it's clear that the above ansatz cannot work, but constant $\psi$ and $\eta$ can solve the equations of motion,
consequently, we have $\log(a(\tau)) = \eta_0(\tau-\tau_0)$. 
After some simple integration, in terms of the cosmic time $t$, the scale factor can be unifyingly written as
\begin{equation}
	a(t) \propto \Big(\eta_0(t-t_0)\Big)^{\frac{4}{3(1+\omega)}},
\end{equation}
which differs significantly from the Einstein's case where the power is $\frac{2}{3(1+\omega)}$ \citep{liddle:2000cg} . 
Hence, our model predicts 
\begin{equation}
	a^{2}H^2 = \Big(\frac{d}{dt}a(t)\Big)^{2} \propto (t-t_0)^{\frac{2(1-3\omega)}{3(1+\omega)}},
\end{equation}
which guarantees the size of the Hubble horizon to be non-decreasing, instead of 
\begin{equation}
	a^{2}H^2 = \Big(\frac{d}{dt}a(t)\Big)^{2} \propto (t-t_0)^{-\frac{2(1+3\omega)}{3(1+\omega)}}
\end{equation}
in Einstein gravity, which causes the flatness problem.  
They also differ greatly in expansion acceleration, as in our model,
\begin{equation}
	a(t)^{-1}\frac{d^2}{dt^2}a(t) = \frac{4}{9(1+\omega)^2} \frac{1-3\omega}{(t-t_0)^2},
\end{equation}
which is non-negative for physical choice of $\omega$ ($0\leq \omega \leq 1/3$).
But in the Einstein's case, the comparative result is 
\begin{equation}
  a(t)^{-1}\frac{d^2}{dt^2}a(t) = \frac{-2}{9(1+\omega)^2}
  \frac{1+3\omega}{(t-t_0)^2},
\end{equation}
which predicts a decelerating expansion of the universe for any physical
$\omega$,
Since there's few observational surveys about the matter and radiation dominant eras, 
we cannot empirically tell which is the correct one.  
Future surveys such as gravitational wave from primordial black holes might be able to help us to rule out one of the candidates.  
Nevertheless, we argue below why the result from the Quadratic Gravity theory with broken conformal symmetry could be correct.

The end of the inflation is usually defined as the end of the first accelerating expansion \citep{Liddle:1999mq}, 
so one would expect that the universe shall have zero expansion acceleration, 
or constant expansion, at that time or even in the reheating era.
However, in Einstein's case, there's a sudden kink to deceleration in the radiation dominant era.  
Comparatively, the radiation dominant era in our model has exactly zero acceleration. 
Furthermore, Einstein gravity also predicts that the deceleration becomes smaller at late time and twist to an exponential expansion dark energy epoch.  
Differently, the acceleration in quadratic gravity becomes larger with time and could smoothly reach the dark energy era.

As a result, it seems that the quadratic gravity can give a smoother expansion history of the universe than Einstein's gravity can.
However, there's one point above we didn't really justify.  
Can the power law expansion in the matter dominant era really smoothly transit
to an exponential expansion dark energy solution?  
In the next section, we will answer this question in the sparse density limit, i.e., in the vacuum.

\subsection{Power Transition Solutions in the Vacuum Limit}

As we mentioned in the end of the last section,
we shall study the expansion dynamics to see whether there are solutions allowing transition from the power law expansion to an exponential one,
but such kind of solutions are difficult to find
in the model with only the perfect fluid.
However, in the early dark energy dominant era, 
the matter and the radiation become sparse due to the expansion,
and one may neglect their presence to study the dynamics in a simpler way.

Motivated by the case of a single perfect fluid, we consider the ansatz
\begin{equation}
	\eta(\tau) = \frac{\eta_0 p(\tau)}{1-\eta_0 \tau}
\end{equation}
and
\begin{equation}
	\psi(\tau) = \frac{\psi_0}{1-\eta_0 \tau},
\end{equation}
since, for single perfect fluid, $q=1$ is independent of the choice of $\omega (0\leq \omega < 1/3)$.  
In the case of single perfect fluid, $p$ indicates the power of the expansion, and $p = 1$ gives the exponential expansion dark energy solution,
accordingly, we would like to find solution that changes $p(\tau)$ from some value larger than $1$ to $1$.

For $k=0$, with our ansatz, the scalar field equation of motion \eqref{eq:scalar_eom} is equivalent to
\begin{equation} \label{eq:trans}
	-\frac{1-\eta_0 \tau}{\eta_0} p'(\tau) = p(\tau)^2 + p(\tau) -  p_0,
\end{equation}
where
\begin{equation}
	p_0 = \frac{2\eta^2_0 - \psi_0^2}{(1+6\xi)\eta^2_0}.
\end{equation}
After substituting this back to \eqref{eq:ansatz}, 
which solves the generalized Einstein equation, 
power matching gives $\gamma = 0$ and the relation
$24(3\alpha+\beta) \eta_0^4 p_0^2 = 2\psi_0^2 \eta_0^2 - \nu \psi_0^4$ between the coefficients,
which subsequently implies
\begin{equation}
	p_0 = \frac{2(1+6\xi)}{(1+6\xi)^2 + 24(3\alpha+\beta)\nu}.
\end{equation}
Hence, $p_0$ is completely determined by the theory parameters.
Provided that $1+4p_0 >0$, then the polynomial $p^2+p-p_0$ has two different real roots, say $p_1 > p_2$,
and the equation \eqref{eq:trans} can be integrated to 
\begin{equation} \label{eq:p_in_tau}
	p(\tau) = \frac{p_1 + (p_1+1) A_0(1-\eta_0 \tau)^{2 p_1+1}}{1-A_0(1-\eta_0 \tau)^{2 p_1+1}}
\end{equation}
for some integration constant $A_0$ and by writing $p_2 = -1 - p_1$.

Suppose that $A_0>0, \eta_0 >0, p_1 = 1$, then for $\tau <
\eta_0^{-1}$, and $p(\tau) >1$ clearly.  Hence, the solution describes
a transition from some power larger than 1 smoothly to 1 at $\tau =
\eta_0^{-1}$, which shall be the conformal time at the end of the
universe.  Also, from this one can see, if we want the desired
transition to happen, then it's necessary that $p_0$ cannot be zero,
consequently, $1+6\xi$ cannot be zero.  As a result, we expect the
coupling between the scalar field and the scalar curvature is not
conformally invariant.  And it's one of the fundamental reasons why
the conformal symmetry of the quadratic gravity theory should be
broken.

\subsection{Numerical Results with Multiple Species of Perfect Fluids}

In this section, we fit the expansion history of the model with the
supernova data.  Difficulties in solving analytic solutions with the
presence of both cold dust and radiation make us resort to numerical
methods, but we shall still take a closer look at the equations of
motion.  Since the supernova data \citep{2018_sharov} are presented in
terms of the redshift $z = a^{-1}-1$, we will rewrite the equations of
motion in terms of the redshift.  Equations \eqref{eq:einstein_eq} and
\eqref{eq:scalar_eom} then respectively take the form
\begin{equation}
\begin{split}
	\frac{d^2\eta}{dz^2} 
		&= -\Big(\frac{1}{2\eta}\frac{d\eta}{dz} + \frac{1}{1+z}\Big)\frac{d\eta}{dz} + \frac{3\eta}{2(1+z)^2} + \frac{k}{(1+z)^2\eta} - \frac{k^2}{2(1+z)^2\eta^3}\\
		& + \frac{1}{48(3\alpha+\beta)(1+z)^2\eta^3}\Big( \nu\psi^4 + 12k\xi\psi^2 - 2(1+6\xi)\Big(2(1+z)\frac{d\psi}{dz} + \psi\Big) \eta^2 \psi \\
		& - 2(1+z)^2\eta^2 \Big(\frac{d\psi}{dz}\Big)^2 + 8 \Big(\tilde{\rho_r}(0)+\frac{\tilde{\rho}_m(0)}{1+z}  \Big)\Big)
\end{split}
\end{equation}
and 
\begin{equation}
\begin{split}
	\frac{d^2\psi}{dz^2}  = &-\Big(\frac{1}{\eta}\frac{d\eta}{dz} + \frac{1}{1+z}\Big)\frac{d\psi}{dz}  \\
		&+ \frac{1}{(1+z)^2}\Big(\frac{6k\xi}{\eta^2} + (1+6\xi)\Big(1-\frac{1+z}{\eta}\frac{d\eta}{dz}\Big)\Big)\psi + \frac{\nu}{(1+z)^2\eta^2} \psi^3.
\end{split}
\end{equation}
For convenience, let us introduce the redefinition 
$\hat{\eta}(z) := \eta(\tau(z))/\eta(0)$
and  $\hat{\psi}(z) := \psi(\tau(z))/\eta(0)$,
which makes $\eta(0)$ the only dimensional quantity.
Next, let us check the redundancy of the parameters.
Since the Hubble parameter $H(z) = (1+z)\eta$ is the only variable to fit,
we shall ``fix'' any parameter transformation that leaves it unchanged.
In fact, we find the Hubble parameter is unchanged under the transformation
$\hat{\psi} \rightarrow \Omega \hat{\psi}, 3\alpha+\beta \rightarrow \Omega^2 (3\alpha +\beta),	
\nu \rightarrow \Omega^{-2} \nu, 			
\xi \rightarrow \xi, \tensor{T}{^\mu_\nu} \rightarrow \Omega^2 \tensor{T}{^\mu_\nu}$, 
for any real number $\Omega$. 
To fix this redundancy, we set $\Omega$ so that $\hat{\psi}(z=0) =1$ 
and still keep the notation without ambiguity.  
As a result, the
eight parameters to fit are the quadratic gravity coefficient
$24(3\alpha+\beta)$, the coupling constants $1+6\xi$ and $\nu$, 
the present value of Hubble parameter $H_0$, the respective present
derivatives of the Hubble parameter and the scalar field
$\frac{d}{dz}\hat{\eta}(0)$ and $\frac{d}{dz}\hat{\psi}(0)$, the
present density of dust $\rho_m(0)$, and the matter-radiation equity
redshift $z_\text{eq}$.

Fitting the numerical solutions with the 57 data points
summarized in \citep{2018_sharov} by minimizing the error 
$\chi^2 = \sum_i \Big( H_{\text{th}}(z_i)-H_{\text{obs}}(z_i) \Big)^2/\sigma(z_i)^2$
gives the following best-fit parameters (Table 1),
plot (Figure 1), and confidence regions (Figure 2).
Here $N$ is the sample size and $n$ is the number of parameters,
AIC is the Akike Information Criteria $\chi^2_\text{min} + 2n$, and
BIC is the Bayesian Information Criteria $\chi^2_\text{min} + n \log N$.

\begin{table}[!hbp]
	\caption{\label{table:fitting3777}Best-Fitting Parameter Values and Statistical Quantities}
	\centering
	\small
	\setlength\tabcolsep{2.5pt}
	\begin{tabular}{c c c c c c c c c c c}
	\hline
	24($\alpha+3\beta$) & 1+6$\xi$  & $\nu$   & $\rho_m(0)$ & $z_{eq}$ & $\frac{d}{dz}\hat{\psi}(0)$ & $H_0$ \footnotemark[1] & $\frac{d}{dz}\hat{\eta}(0)$ & $\chi^2_\text{min}$ & AIC & BIC \\
	7.7747              & 3.4545    & -0.5996 &  6.3583$\cross 10^{-5}$    &  3675   & -2.3898                    & 70.8040               & -0.7618    & 33.0481  & 49.0481  & 49.9510 \\
	\end{tabular}
\end{table}
\footnotetext[1]{$H_0$ is the present value of the Hubble parameter, and the unit of both $H_0$ and $\eta_0$ is km s$^{-1}$Mpc$^{-1}$. } 

\newpage

\begin{figure}[htp!]
	\centering
	\includegraphics[width = 0.78\linewidth]{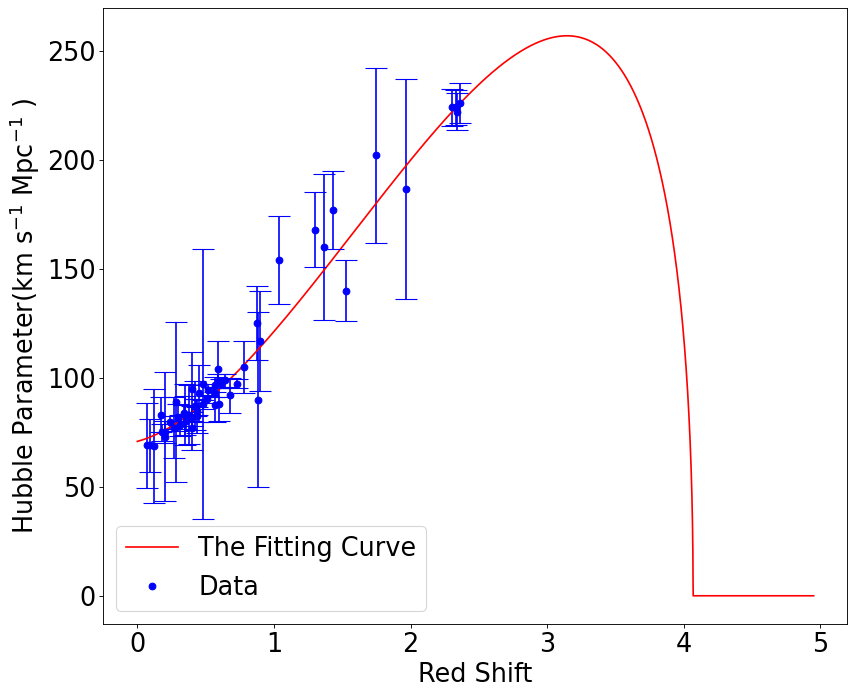}
	\caption{\label{fig:3675}The Best Fit Curve with 57 Points Hubber Parameter Data}
	\vspace{1ex}
\end{figure}

\begin{figure}[!htp]
	\centering
	\includegraphics[width=\linewidth]{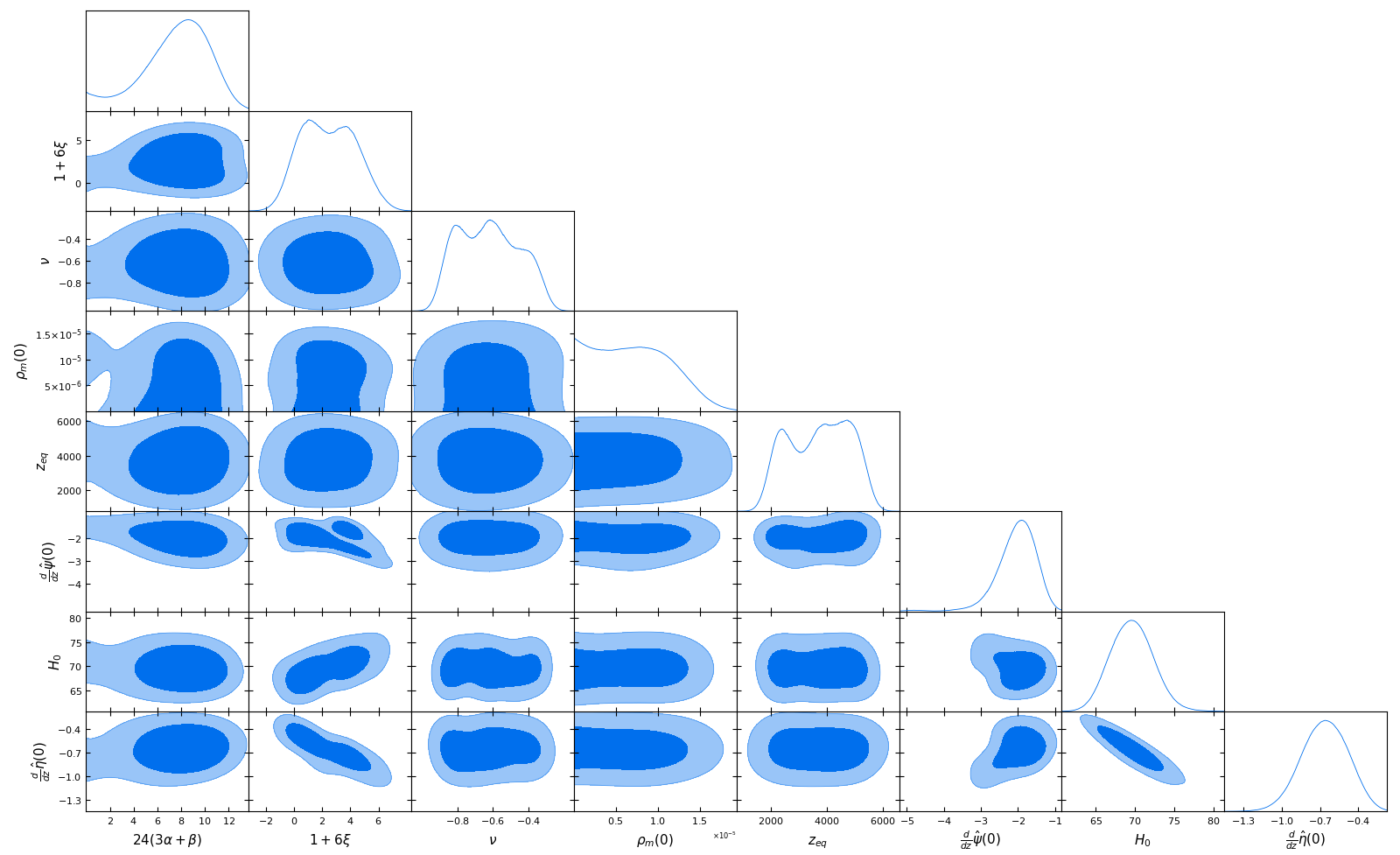}
	\caption{\label{fig:ndist}One and Two $\sigma$ Confidence Regions Of The Fit}
	\vspace{1ex}
\end{figure}

\newpage 

The very first thing one may observe in Fig. \ref{fig:3675} is that
the best-fit Hubble parameter drops to zero at around redshift
$z\approx 4$, which is certainly unphysical since the recombination
redshift $z_\text{re} \approx 1100$ is model independent and we do
observe its aftermath, the cosmological microwave background. Also, we
do have compact object with redshift larger than $5$
\citep{Kroupa_2020}.  In the next section, we will numerically resolve
this problem of ``recombination safety''.

The best-fit solution gives $\chi^2$ error $33.0481$. 
Hence, the root mean square error (RMSE) is around $0.761$, which is still relatively acceptable.
Also, the Hubble parameter falls between the predictions of $\Lambda$-CDM by supernova data and the cosmological microwave background power spectrum,
which are respectively around $72$ km s$^{-1}$ Mpc$^{-1}$ and $67$ km s$^{-1}$ Mpc$^{-1}$ \citep{Di_Valentino_2021}.
Consequently, it's possible that the quadratic gravity could also resolve the Hubble tension issue.
Furthermore, quadratic gravity gives the matter-radiation equity redshift $z_{eq} = 3675$, 
which does not change a lot from the $\Lambda$-CDM prediction $z_{eq} = 3379 \pm 22$ from the 2018 Planck data \citep{2020_planck}.

In Fig. \ref{fig:ndist}, 
we implement the GetDist package and Monte-Carlo sampling to study the confidence regions.
The diagonal plots are the probabilities of each parameters calculated by the Kernel Density Estimation (KDE) method,
and the off diagonal plots are the joint probabilities of each choice of two parameters.
Multiple peaks can be seen in the probabilities plot of $24(3\alpha+\beta), 1+6\xi, \nu, \rho_m(0),$ and $z_{eq}$,
which means there are overfitting from these parameters. 
But since we do have several parameters in this theory, 
it’s not a surprising result, thought some redundancy has been previously removed by us.
However, it's surprising that $\frac{d}{dz}\hat{\psi}(0)$ has a single peak with an order one 1-$\sigma$ region.

In order to guarantee the ``recombination safety'', we come up with the following backward fitting method,
which imposes a ``survival condition'' to the numerical solutions. 

\subsection{The Backward Fitting Method}

In the last section, we show that the numerical solution fitting the type
Ia supernovae best freezes to zero at around redshift $z\approx 4$,
and this behavior is certainly undesired and unphysical
due to the existence of the CMB,
which was generated at the end of the recombination, when the redshift was $z\approx 1100$.
To find numerical solutions which can survive to $z\approx 1100$
without increasing much the $\chi^2$ error, we propose to choose some
$z_{par}>0$ and set some initial values at $z_{par}$ to solve the
initial value problem from $z_{par}$ back to $z = 0$.  Consequently,
we can treat the initial values at $z_{par}$ as parameters to fit the
supernovae data, and get the best-fit parameters for each choice of
$z_{par}$. In this way, we can make the Hubble parameter positive at
least from the redshift $z = 0$ to the chosen $z_{par}$.  After the
fitting, we then extrapolate the numerical solutions and check  whether
it can survive to the recombination redshift or not.

As a result, asides from
the quadratic gravity coefficient $24(3\alpha+\beta)$, 
the coupling constants $1+6\xi$ and $\nu$,
the present density of dust $\rho_m(0)$, 
and matter-radiation equity redshift $z_\text{eq}$,
which shall be independent of $z_{par}$. 
We also have three parameters which are $z_{par}$ dependent, which are 
the Hubble parameter at $z_{par}, H_{par}$,
the respective derivatives of the Hubble parameter and the scalar field at $z_{par}$, $\frac{d}{dz}\hat{\eta}(z_{par})$ and $\frac{d}{dz}\hat{\psi}(z_{par})$.
We provide the plots with different choices of $z_{par}$ below.
Also, we provide the best-fit parameters and their corresponding $\chi^2$ errors for different choices of $z_{par}$. 

\newpage
\vspace{1ex}
\begin{figure}[htp!]
	\centering
	\includegraphics[width = 0.9\linewidth]{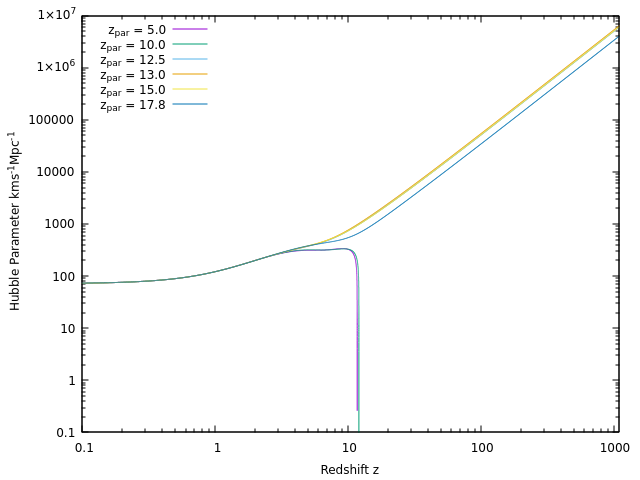}
	\caption{\label{fig:zmaxdiff} The Backward Best-Fit Curves with Different Partition Redshifts}
	\vspace{4ex}
\end{figure}

\begin{table}[!htbp] 
	\centering
	\caption{The Best-Fit Parameters for each $z_\text{par}$}
	\label{table:best_diff_zpar}
	\resizebox{\columnwidth}{!}{
	\setlength\tabcolsep{2.5pt}
\setstretch{1.15} 
	\begin{tabular}{c c c c c c c c c c}
		\\[-1.8ex]\hline
		\hline \\[-1.8ex]
$z_\text{par}$ &24($3\alpha+\beta$)&1+6$\xi$ & $\nu$ & $\rho_m(0)$ & $z_{eq}$ & $\frac{d}{d\zeta}\hat{\psi}(z_\text{par})$ & $H_\text{par}$ & $\frac{d}{d\zeta}\hat{\eta}(z_\text{par})$&$\chi^2 $ \\
		\hline \\[-1.8ex]
	5.0  & $1.54\times10^1$ & $2.44\times10^0$ & $-2.62\times10^0$ & $2.59\times10^0$    & $3675.0$ & $1.91\times10^0$ & $3.87\times10^1$ & $2.87\times10^{-1}$    &35.1222\\
	5.3  & $2.15\times10^0$ & $3.57\times10^0$ & $-1.42\times10^1$ & $1.48\times10^{-5}$ & $3675.0$ & $2.46\times10^{-1}$ & $4.52\times10^1$ & $2.43\times10^{-1}$ &33.0556\\
	5.8  & $2.92\times10^0$ & $4.00\times10^0$ & $-1.24\times10^1$ & $3.00\times10^{-6}$ & $3675.0$ & $3.28\times10^{-1}$ & $4.31\times10^1$ & $1.84\times10^{-1}$ &33.0533\\
	6.5  & $4.08\times10^0$ & $3.80\times10^0$ & $-8.31\times10^0$ & $2.39\times10^{-5}$ & $3675.0$ & $4.38\times10^{-1}$ & $4.04\times10^1$ & $1.34\times10^{-1}$ &33.0548\\
	7.2  & $6.21\times10^0$ & $3.61\times10^0$ & $-5.05\times10^0$ & $7.23\times10^{-6}$ & $3675.0$ & $5.37\times10^{-1}$ & $3.88\times10^1$ & $9.76\times10^{-2}$ &33.0575\\
	8.0  & $1.56\times10^1$ & $3.72\times10^0$ & $-2.04\times10^0$ & $7.10\times10^{-6}$ & $3675.0$ & $1.03\times10^0   $ & $3.24\times10^1$ & $1.23\times10^{-1}$ &33.0577\\
	9.0  & $2.19\times10^1$ & $2.97\times10^0$ & $-1.01\times10^0$ & $6.47\times10^0   $ & $3674.9$ & $1.10\times10^0   $ & $3.31\times10^1$ & $1.04\times10^{-1}$ &33.0694\\
	10.0 & $4.98\times10^1$ & $2.78\times10^0$ & $-3.99\times10^{-1}$ & $1.01\times10^1$ & $3676.8$ & $7.65\times10^{-1}$ & $4.70\times10^1$ & $-1.27\times10^{-2}$ &33.0727\\
	11.0 & $1.58\times10^2$ & $2.93\times10^0$ & $-1.46\times10^{-1}$ & $6.16\times10^0$ & $3673.5$ & $5.63\times10^{-1}$ & $7.01\times10^1$ & $-5.75\times10^{-2}$ &33.0747\\
	12.0 & $1.69\times10^2$ & $2.59\times10^0$ & $-1.08\times10^{-1}$ & $7.95\times10^0$ & $3673.6$ & $4.93\times10^{-1}$ & $7.39\times10^1$ & $-5.60\times10^{-2}$ &33.0810\\
	13.0 & $2.49\times10^2$ & $2.47\times10^0$ & $-6.72\times10^{-2}$ & $8.84\times10^0$ & $3673.7$ & $4.52\times10^{-1}$ & $8.22\times10^1$ & $-5.69\times10^{-2}$ &33.0842\\
	14.0 & $2.40\times10^2$ & $2.22\times10^0$ & $-5.51\times10^{-2}$ & $1.23\times10^1$ & $3674.8$ & $4.49\times10^{-1}$ & $7.92\times10^1$ & $-5.18\times10^{-2}$ &33.0888\\
	15.0 & $2.41\times10^2$ & $2.19\times10^0$ & $-5.22\times10^{-2}$ & $9.25\times10^0$ & $3674.7$ & $4.47\times10^{-1}$ & $8.67\times10^1$ & $-5.19\times10^{-2}$ &33.0910\\
	16.0 & $2.42\times10^2$ & $1.96\times10^0$ & $-4.29\times10^{-2}$ & $9.36\times10^0$ & $3674.6$ & $3.16\times10^{-1}$ & $8.95\times10^1$ & $-4.93\times10^{-2}$ &33.0939\\
	17.0 & $2.28\times10^2$ & $2.35\times10^0$ & $-5.58\times10^{-2}$ & $4.81\times10^0$ & $3674.5$ & $1.98\times10^{-1}$ & $1.03\times10^2$ & $-5.02\times10^{-2}$ &33.0978\\
	17.8 & $2.51\times10^2$ & $8.35\times10^0$ & $-3.45\times10^{-1}$ & $5.70\times10^0$ & $3674.5$ & $2.20\times10^{-1}$ & $6.62\times10^1$ & $-4.70\times10^{-2}$ &33.0709\\
	\end{tabular}}
\setstretch{1.0} 
\end{table}
\newpage 

In Fig. \ref{fig:zmaxdiff}, one see that the Hubble parameters still
drop to zero at around $z\approx 10$ with choices $z_{par} = 5.0,
10.0,$ or $12.5$.  However, if we choose $z_{par} \geq 13.0$, then the
Hubble parameter will be positive from redshift $z=0$ to $1100$, which
gives a physical expansion history instead of a finite one.  In fact,
we find that numerical solutions with these $z_{par}$ can be
extrapolated to at least $z=1500$, but we didn't study the maximal
redshifts to which these numerical solutions can be extended since
there are many other important effects which should be considered
before the end of the recombination era, such as the baryon acoustic
oscillations (BAO).

We note that this backward fitting method only increases the $\chi^2$
errors by around $0.1\%$ from the normal fitting method ($\chi^2 =
33.0481$).  Hence, the backward fitting method gives best-fit
numerical solutions as good as that given by the normal fitting
method.  This ``survival phase transition'' parameterized by $z_{par}$
could arise from the competition between the best-fit dropping and
surviving solutions.  For each given $z_{par} < 13.0$, the best-fit
solution in the dropping family might have smaller $\chi^2$ error than
that in the surviving family, but the $\chi^2$ error of the best-fit
curve in the surviving family might become smaller than that in the
dropping family after $z_{par} = 13.0$, which makes the survival ones
the overall best-fit numerical solutions.
We also note that in Table \ref{table:best_diff_zpar}, 
one  find that some of the parameters which should be $z_{par}$ independent
vary greatly with it.
This might arise from the overfitting nature of this model that we cannot
solve, 
that is, a wide range of parameters can give fairly good fitting,
and a small change of $z_{par}$ could give a large change of the best-fit
parameters.

On the contrary, all the choices of $z_{par}$ we study give $z_{eq} \approx 3675$ and predict almost the same matter-radiation equality redshift. 
It could be because that the overfitting nature of this model could give a good fitting with any choice of $z_{eq}$, 
but we didn't study this systematically.
Last but not least, we note that our best-fit surviving solutions predict a 
typical value for the radius of the Hubble horizon at around a few smaller than $9.8 \times 10^5$ light years when the recombination ended. This is
basically in the same order as the
characteristic size of around $4 \times 10^5$ light years \citep{Jackson_2007}
of the observable universe at the
recombination redshift. Hence, it's unlikely that the expansion history of
the universe after the recombination would be drastically changed.

\section{Conclusion}

A good
theory that aims to resolve the unsolved problems of Einstein's gravity
should
pass all the empirical tests passed by Einstein's gravity.
It is better if it can resolve
some or all
of unsolved problems in a unifying manner.
In this paper, we have considered quadratic gravity theories as such 
candidates.
Quadratic gravity theories classically allow all the Einstein manifolds. 
Hence, it can predict exactly the same results of Mercury precession and
light deflection by massive object
as in the
Einstein's gravity.
Moreover, 
quadratic gravity theories are known to be renormalizable and able to
trigger sufficient inflation, although the essential difficulty of
quantum gravity might not be merely the renormalizability but the
definition of time when the metric fluctuates
quantum mechanically.
Furthermore,
quadratic gravity theories could explain dark matter, as conformal
gravity is a special case, despite the necessity that we shall still
find similar solution that fits the galaxy rotation curve with the
conformal symmetry broken.

Consequently,
it is interesting to study whether quadratic gravity theories
can lead to an accelerating expanding universe, which is the central
goal of this paper.  In order to
analyze this possibility, we study the
analytic properties of the quadratic gravity theories with broken
conformal symmetry, including the integrability, in the first part of
this paper.  Three sufficient conditions to the complete
integrability are listed.
We show that this integrability could
be an accidental one as we do not know of any symmetry that would
give rises to this integrability.

In the second part of this paper, we analytically study the expansion
behavior of the quadratic gravity theories with broken conformal
symmetry due to the presence of a single perfect fluid.
We find that the resulting behaviour  differs
largely from that of Einstein's gravity.  In the sparse density limit,
we also show that the single perfect fluid solution can smoothly
transit to the dark energy solution.  Hence,
quite interestingly,
a smooth cosmological
history from the start of the inflation to the present could be
described by a single model.  We show numerically that quadratic
gravity theories with broken conformal symmetry can fit the type Ia
supernova data well with the presence of both cold dust and radiation,
despite the fact that the backward fitting method is used to conquer
some numerical problem and that fitting to the CMB power spectrum
shall also be done.  Nevertheless, about whether quadratic gravity
theories can explain the dark energy or not, the answer is likely
positive.

Summarizing, although
quadratic gravity theories might be merely an effective
theory of a more unifying theory, which should take the gauge fields,
fermions, and the Higgs field all into the consideration,
they might be useful to constrain the unifying theory and helpful to
find the realistic vacuum in the moduli space of a candidate unifying theory.
Furthermore, we could also rule out the possibility that gravity is non-quantum \citep{Boughn_2009}
if an effective theory of some quantum gravity theory can describe our universe, 
especially when we find the evidence that the parameters of this effective theory are running. 

\section*{Acknowledgements}

We thank Chong-Sun Chu and Siye Wu for useful discussions and comments. This
paper is partly the author’s master work under Chong-Sun Chu’s supervision.

\newpage
\bibliographystyle{unsrt}
\bibliography{references}

\end{document}